\documentclass[12pt]{article}

\usepackage{amsmath,amssymb}
\usepackage{bm}
\usepackage{graphicx, color}
\begin{document}
\title{
\begin{flushright}
\ \\*[-80pt] 
\begin{minipage}{0.2\linewidth}
\normalsize
KUNS-2151 \\*[50pt]
\end{minipage}
\end{flushright}
{\Large \bf 
Soft supersymmetry breaking terms \\ from 
$A_4$ lepton flavor symmetry 
\\*[20pt]}}

\author{
\centerline{
Hajime~Ishimori$^{1,}$\footnote{E-mail address: ishimori@muse.sc.niigata-u.ac.jp},   
~Tatsuo~Kobayashi$^{2,}$\footnote{E-mail address: kobayash@gauge.scphys.kyoto-u.ac.jp}, } \\ 
\centerline{
~Yuji~Omura$^{3,}$\footnote{E-mail address: omura@scphys.kyoto-u.ac.jp}, 
Morimitsu~Tanimoto$^{4,}$\footnote{E-mail address: tanimoto@muse.sc.niigata-u.ac.jp} }
\\*[20pt]
\centerline{
\begin{minipage}{\linewidth}
\begin{center}
$^1${\it \normalsize
Graduate~School~of~Science~and~Technology,~Niigata~University, \\ 
Niigata~950-2181,~Japan } \\
$^2${\it \normalsize 
Department of Physics, Kyoto University, 
Kyoto 606-8502, Japan} \\
$^3${\it \normalsize 
Department of Physics, Kyoto University, 
Kyoto 606-8501, Japan} \\
$^4${\it \normalsize
Department of Physics, Niigata University,~Niigata 950-2181, Japan } 
\end{center}
\end{minipage}}
\\*[50pt]}

\date{
\centerline{\small \bf Abstract}
\begin{minipage}{0.9\linewidth}
\medskip 
\medskip 
\small
We study the supersymmetric model with the $A_4$ 
lepton flavor symmetry, in particular 
soft supersymmetry breaking terms, which are predicted from 
the $A_4$ lepton flavor symmetry.
We evaluate soft slepton masses and A-terms within 
the framework of supergravity theory.
Constraints due to experiments of 
flavor changing neutral current processes are examined.     
\end{minipage}
}

\begin{titlepage}
\maketitle
\thispagestyle{empty}
\end{titlepage}

\section{Introduction}


Recent experiments of the neutrino oscillation go into the new  phase  
of precise  determination of
 mixing angles and mass squared  differences \cite{maltoni,fogli}. 
Those indicate the  tri-bimaximal mixing  for three flavors 
 in the lepton sector \cite{HPS}.
Indeed, various types of models leading to the tri-bimaximal mixing 
have been proposed, e.g. by assuming 
several types of non-Abelian flavor symmetries.

One of natural models realizing the tri-bimaximal mixing  
has been proposed  
based on the non-Abelian finite group $A_4$.
Since the original papers \cite{A4} on the application of the 
non-Abelian discrete symmetry $A_4$ to quark and lepton families, 
much progress has been made in understanding the  tri-bimaximal 
mixing for neutrinos in a number of specific models 
\cite{Alta1,Alta2,A41,kaneko}.
Therefore, it is important to 
clarify the physical implication of the $A_4$ model carefully.

The supersymmetric extension of the standard model is 
one of interesting candidates for physics beyond the 
weak scale.
Within the framework of supersymmetric models, 
flavor symmetries constrain not only quark and 
lepton mass matrices, but also mass matrices of 
their superpartners, i.e., squarks and sleptons.
That is, flavor symmetries realizing realistic quark/lepton 
mass matrices would lead to specific patterns of 
squark and slepton mass matrices as their predictions, 
which could be tested in future experiments.
For example, $D_4$ flavor models \cite{Grimus,D4,D4-2,Ishimori:2008gp}
would also lead to the lepton tri-bimaximal mixing.
Their supersymmetric models have been studied in
Ref.~\cite{Ishimori:2008ns} and it is shown that 
the $D_4$ flavor models predict the degeneracy between 
the second and third families of slepton masses.\footnote{
The $D_4$ flavor symmetry can be realized in superstring 
models~\cite{Kobayashi:2004ya,Kobayashi:2006wq}. From 
this viewpoint, the $D_4$ flavor symmetry as well as 
certain flavor symmetries are interesting, too.}
The $A_4$ model would lead to a different prediction in 
slepton masses.

Although squarks and sleptons have not been detected yet, 
their mass matrices are strongly constrained by 
experiments of flavor changing neutral current (FCNC) 
processes~\cite{FCNCbound}.\footnote{See also 
e.g  Ref.~\cite{Chankowski:2005jh} and references therein.}
When squark and slepton masses are of order of the weak scale, 
the FCNC experimental bounds, in particular the 
$\mu \rightarrow e \gamma$ decay, requires 
strong suppression of off-diagonal elements in squark and slepton 
mass squared matrices in the basis, where 
fermion mass matrices are diagonalized.
Non-Abelian flavor symmetries and certain types of 
their breaking patterns are useful to suppress FCNCs.
(See e.g. \cite{Kobayashi:2003fh,Ko:2007dz,Ishimori:2008ns}.) 
In addition to flavor symmetries, their breaking 
patterns are important to derive quark and lepton 
mass matrices and to predict squark and slepton mass 
matrices.
Thus, it is important to study which pattern of slepton mass 
matrices is predicted from the $A_4$ model and 
to examine whether the predicted pattern of slepton mass 
matrices is consistent with the current FCNC experimental 
bounds.
That is the purpose of this paper.

The paper is organized as follows.
In Section 2, we review the $A_4$ model~\cite{Alta2}, 
showing values of parameters consistent with
neutrino oscillation experiments.
In Section 3, we evaluate soft 
supersymmetry (SUSY) breaking terms of 
sleptons, i.e. soft scalar mass matrices and A-terms.
We examine FCNC constraints on those SUSY breaking terms 
as mass insertion parameters.
Section 4 is devoted to conclusion and discussion.
In Appendix, we give a brief review on the $A_4$ group.

\section{$A_4\times Z_3$ model for leptons}

Here, we discuss the $A_4$ model~\cite{Alta2} leading to 
the tri-bimaximal mixing and show proper values of parameters. 
In the non-Abelian finite group $A_4$, 
there  are twelve group elements and
 four irreducible representations: $\bf 1$, $\bf 1'$, $\bf 1''$ and  $\bf 3$.
We consider the supersymmetric $A_4$ model based on \cite{Alta2},
with the $A_4$ and $Z_3$ charge assignments listed in Table \ref{a4}. 
Under the $A_4$ symmetry, the chiral superfields for 
three families of the left-handed lepton doublets  $L_I$ 
$(I=e,\mu,\tau)$
are assumed to transform as  $\bf 3$, while  
the right-handed ones of the  charge lepton singlets 
$R_e$,  $R_\mu$ and  $R_\tau$ are assigned with   
$\bf 1$, $\bf 1'$, $\bf 1''$, respectively.
The third row of Table \ref{a4} shows how each chiral multiplet transforms 
under $Z_3$, where $\omega = e^{2\pi i/3}$. 
The flavor symmetry is spontaneously broken 
by vacuum expectation values (VEV) of 
two  ${\bf 3}'s$, $\chi_i$, $\chi'_i$, and by one  singlet,
 $\chi(\bf 1)$, which are   $SU(2)_L\times U(1)_Y$ singlets.
Their $Z_3$ charges are also shown in Table \ref{a4}. 
Here and hereafter, we follow the convention that 
the chiral superfield and its lowest component are 
denoted by the same letter.

\begin{table}[t]
\begin{center}
\begin{tabular}{|c|cccc||cc||ccc|c|}
\hline
              &$(L_e,L_\mu,L_\tau)$ & $R_e$  & $R_\mu $ &
              $R_\tau $ &  $H_u$ & $H_d$  & $(\chi_1,\chi_2,\chi_3)$
              &$(\chi'_1,\chi'_2,\chi'_3)$ &$\chi$ & $\Phi$  \\ \hline
$A_4$      &{\bf 3}      & ${\bf 1}$            & ${\bf 1}'$   & ${\bf
  1}''$      &    {\bf 1}&    {\bf 1}   & {\bf 3}         &{\bf 3}&{\bf 1} &{\bf 1}     \\
$Z_3$      &  $\omega$      &  $\omega^2$         & $\omega^2$ &
$\omega^2$      & 1   &     1 &     1                       &$\omega$
&$\omega$  & 1 \\ 
$U(1)_F$ & 0 & $2q$ & $q$ & 0 &  0 & 0 & 0 & 0 & 0 & $-1$ \\
\hline
\end{tabular}
\caption{$A_4$, $Z_3$ and $U(1)_F$ charges for leptons and scalars\label{a4}}
\end{center}
\end{table}

The allowed terms in the superpotential including 
charged leptons are written by 
\begin{eqnarray}
W_L
&=&\frac{y_e}{\Lambda}( L_e\chi_1+ L_\mu\chi_3+ L_\tau\chi_2)
R_e H_d
+\frac{y_\mu}{\Lambda}( L_e\chi_2+ L_\mu\chi_1+ L_\tau\chi_3)
R_\mu H_d
\nonumber\\
&&+\frac{y_\tau}{\Lambda}( L_e\chi_3+ L_\mu\chi_2+ L_\tau\chi_1)
R_\tau H_d \ ,
\end{eqnarray}
where $y_e$, $y_\mu$ and $y_\tau$ are couplings, 
and $\Lambda$ is the cut-off scale of the effective superpotential. 
In order to obtain
the natural hierarchy among lepton masses $m_e$, $m_\mu$ and $m_\tau$,
the  Froggatt-Nielsen mechanism \cite{FN} is  introduced as 
 an additional $U(1)_F$ flavor symmetry under which
only the right-handed lepton sector is charged.
The $U(1)_F$ charge values are taken as $0, \ q$ and $2q$ for
$R_\tau$, $R_\mu$ and  $R_e$, respectively.
By assuming that the flavon $\Phi$, 
carrying a negative unit charge of $U(1)_F$,
 acquires a VEV 
$\langle \Phi \rangle/\Lambda\equiv \lambda<1$,
the following magnitudes of couplings are realized through 
the Froggatt-Nielsen mechanism, 
\begin{eqnarray}
 y_\tau\simeq {\cal O}(1), \qquad y_\mu\simeq {\cal O}(\lambda^q)  , \qquad
y_e\simeq {\cal O}(\lambda^{2q}) \ .
\end{eqnarray}
When $q = 1$, we estimate $\lambda \sim 0.02$. 
The $U(1)_F$ charges are shown in the fourth row of Table~\ref{a4}.

The superpotential of the neutrino sector is given as 
\begin{eqnarray}
W_\nu
&=&\frac{y_1}{\Lambda^2}(L_e L_e+ L_\mu L_\tau+ L_\tau L_\mu)H_uH_u\chi
\nonumber\\&&
+\frac{y_2}{3\Lambda^2}
[ (2L_e L_e- L_\mu L_\tau- L_\tau L_\mu)\chi_1'
+(- L_e L_\tau +2 L_\mu L_\mu-  L_\tau L_e)\chi_2'
\nonumber\\&&
+(- L_e L_\mu- L_\mu L_e+2 L_\tau L_\tau)\chi_3' ] H_uH_u \ ,
\end{eqnarray}
where $y_1$ and $y_2$ are couplings  of ${\cal O}(1)$.
After the $A_4\times Z_3$ symmetry is spontaneously broken 
by VEVs of $\chi_i$, $\chi'_i$ and $\chi$,  
the charged lepton mass matrix $M_l$ and the neutrino mass 
matrix $M_\nu$ are obtained as follows,
\begin{eqnarray}
M_l
&=&\frac{v_d}{\Lambda} 
\left(
  \begin{array}{ccc}
y_e \langle \chi_1\rangle  &  y_e \langle \chi_3\rangle &   
y_e \langle \chi_2\rangle \\ 
 y_\mu \langle \chi_2\rangle  &  y_\mu \langle \chi_1\rangle &  
y_\mu \langle \chi_3\rangle \\ 
 y_\tau \langle \chi_3\rangle & y_\tau \langle \chi_2\rangle& 
 y_\tau\langle \chi_1\rangle\\ 
\end{array} \right) \ ,
\label{charged}
\end{eqnarray}
and
\begin{eqnarray}
M_\nu
&=& \frac{v_u^2}{3\Lambda^2}
\left(
  \begin{array}{ccc}
3y_1  \langle \chi\rangle +2y_2 \langle \chi'_1\rangle   
&  -y_2 \langle \chi'_3\rangle  &  -y_2 \langle \chi'_2\rangle \\ 
-y_2 \langle \chi'_3\rangle   &  2y_2 \langle \chi'_2\rangle  
& 3y_1 \langle \chi\rangle -y_2 \langle \chi'_1\rangle  \\ 
-y_2 \langle \chi'_2\rangle    
& 3y_1 \langle \chi\rangle -y_2 \langle \chi'_1\rangle  
 & 2y_2 \langle \chi'_3\rangle    \\ 
\end{array} \right),
\label{neutrino}
\end{eqnarray}
where  $v_u$ and $v_d$ denote 
VEVs of Higgs doublets, i.e.  
$\left< H_u\right>=v_u$ and $\left< H_d\right>=v_d$.
Furthermore, we define $\tan \beta = v_u/v_d$.

If one can take  the VEVs of gauge singlet
scalar fields $\chi$, $\chi_i$ and $\chi'_i$ as follows 
\begin{eqnarray}
\left< \chi\right>=V, \quad
\left< (\chi_1,\chi_2,\chi_3) \right>=(V_l,0,0), \quad
\left< (\chi'_1,\chi'_2,\chi'_3) \right>=(V_\nu,V_\nu,V_\nu),
\label{vev}
\end{eqnarray}
which are  actually
  one of minima in the scalar potential at the leading order as 
shown in ref.\cite{Alta2},
the mass matrices of charged leptons and neutrinos are reduced  to
\begin{eqnarray}
M_l
&=&\frac{v_d V_l}{\Lambda} 
\left(
  \begin{array}{ccc}
y_e   &  0 &  0 \\ 
0  &  y_\mu & 0 \\ 
0  & 0 & y_\tau \\ 
\end{array} \right),
\label{ME}\\
M_\nu
&=& \frac{v_u^2}{3\Lambda^2}
\left(
  \begin{array}{ccc}
3y_1V+2y_2V_\nu   &  -y_2V_\nu &  -y_2V_\nu \\ 
-y_2V_\nu   &  2y_2V_\nu & 3y_1V-y_2V_\nu  \\ 
-y_2V_\nu   & 3y_1V-y_2V_\nu   & 2y_2V_\nu   \\ 
\end{array} \right).
\end{eqnarray}
The charged lepton mass matrix is diagonal.
The neutrino mass matrix can be simplified as
\begin{eqnarray}
M_\nu
= \frac{y_1v_u^2V}{\Lambda^2}
\left(
  \begin{array}{ccc}
1  &  0 &  0 \\ 
0   &  0 & 1 \\ 
0   & 1   & 0   \\ 
\end{array} \right)
+ \frac{y_2v_u^2V_\nu}{3\Lambda^2}
\left(
  \begin{array}{ccc}
2   &  -1 &  -1 \\ 
-1   &  2 & -1  \\ 
-1   & -1   & 2    \\ 
\end{array} \right).
\end{eqnarray}
Then, it is easy to find the tri-bimaximal mixing for the lepton flavor mixing matrix $U_{MNS}$ \cite{MNS} as,
\begin{eqnarray}
U_{MNS}=
\left(
  \begin{array}{ccc}
\sqrt{2/3}   &  \sqrt{1/3} &  0 \\ 
-\sqrt{1/6}   & \sqrt{1/3} & -\sqrt{1/2}  \\ 
-\sqrt{1/6}  & \sqrt{1/3}   & \sqrt{1/2}   \\ 
\end{array} \right).
\end{eqnarray}
On the other hand,  neutrino masses are given by
\begin{eqnarray}
m_\nu= \frac{v_u^2}{\Lambda^2}
(y_1V+y_2V_\nu,y_1V,-y_1V+y_2V_\nu),
\end{eqnarray}
and using the parameter $r=y_2V_\nu/y_1V$
the neutrino masses are expressed as
\begin{eqnarray}
m_\nu= \frac{y_1Vv_u^2}{\Lambda^2}
(1+r,1,-1+r).
\end{eqnarray}
Then, the differences between masses squared  are evaluated as,
\begin{eqnarray}
\Delta m_{\rm atm}^2
=\left |-4r\frac{y_1^2V^2v_u^4}{\Lambda^4} \right |, \qquad
\Delta m_{\rm sol}^2
=r(r+2)\frac{y_1^2V^2v_u^4}{\Lambda^4},
\label{mass2}
\end{eqnarray}
that is,
\begin{eqnarray}
\frac{\Delta m_{\rm atm}^2}{\Delta m_{\rm sol}^2}
=\left |\frac{-4}{r+2}\right |\ ,
\label{ratio}
\end{eqnarray}
which is reconciled with the experimental data 
for $r\sim -1.9$ or $r\sim -2.1$.

Let us estimate numerical values of
 $\alpha_l$, $\alpha_\nu$ and $\alpha$, which  are determined by putting
the neutrino experimental data.
By using Eqs.(\ref{ME}), (\ref{mass2}) and (\ref{ratio}),
 we obtain the following relations,
\begin{eqnarray}
\alpha_l=\frac{m_\tau}{y_\tau v_d},
\quad
\alpha^2=\left |-\frac{\Delta m_{\rm atm}^2\Lambda^2}{4r y_1^2 v_u^4}\right |,
\quad
\frac{y_2\alpha_\nu}{y_1\alpha}
 =-2\pm 4\frac{\Delta m^2_{\rm sol}}{\Delta m^2_{\rm atm}},
\end{eqnarray}
where
\begin{eqnarray}
\alpha_l
=\frac{V_l}{\Lambda},
\qquad
\alpha_\nu
=\frac{V_\nu}{\Lambda},
\qquad
\alpha
=\frac{V}{\Lambda}.
\end{eqnarray}
We use the experimental data, 
\begin{equation}
m_\tau \simeq  1.8{\rm GeV}, \qquad 
\Delta m_{\rm atm}^2 \simeq  2.4\times10^{-3}{\rm eV}^2,
\qquad
\Delta m_{\rm sol}^2 \simeq  7.6\times10^{-5}{\rm eV}^2.
\end{equation}
For example, in the case with $\tan \beta = 3$ and 
$ |y_\tau|\simeq |y_1|\simeq |y_2|\simeq 1$, 
we can estimate 
\begin{eqnarray}
\alpha_l\sim0.03,
\quad
\alpha
\sim 6\times10^{-16}\times
\frac{\Lambda}{1{\rm GeV}},
\quad
\alpha_\nu\simeq 2 \alpha \ .
\end{eqnarray}
Moreover, if the breaking scales are assumed to be approximately 
the same $V_l\sim V_\nu\sim V$, 
the scale $\Lambda\simeq 10^{14} {\rm GeV}$ is  determined. 
As the value of $\tan \beta$ becomes larger, $\alpha_l$ increases.
For example, for $\tan \beta =30$ and $ |y_\tau|\simeq 1$,
we obtain $\alpha_l \sim 0.3$.
Similarly, $\alpha_l$ increases as $|y_\tau|$ decreases.
Thus, the above value $\alpha_l \sim 0.03$ is the smallest value 
for $ |y_\tau| \leq {\cal O}( 1)$.
On the other hand, if we allow a large value of $|y_\tau|$ 
like $|y_\tau| \sim 4 \pi$, the value of $\alpha_l$ 
would be estimated as $\alpha_l \sim 0.002$.
Hereafter, we restrict ourselves to the case with 
$\alpha_l \sim \alpha_\nu \sim \alpha$ and we denote 
their magnitudes by $\tilde \alpha$.


In the above scenario, it is crucial to choose the proper VEVs, i.e., 
(\ref{vev}).
Indeed, such VEVs can be realized by a certain form of superpotential 
at the leading order, as shown in ref.~\cite{Alta2}.
Also, in ref.~\cite{Alta2} it has been shown that when 
the next leading  terms are taken into account,  
VEVs shift in  the order of
 $\langle \chi_i\rangle/\Lambda$ and  $\langle \chi'_i\rangle/\Lambda$.
 Actually, one can obtain
\begin{eqnarray}
&&\left<(\chi_1,\chi_2,\chi_3)\right>
=(1+g_{l_1}\tilde \alpha,~ g_{l_2}\tilde \alpha,~g_{l_3}\tilde
\alpha)V_l, 
\nonumber\\
&&\left<(\chi_1',\chi_2',\chi_3')\right>
=(V_\nu+g_{\nu_1}\tilde \alpha,~V_\nu+g_{\nu_2}\tilde \alpha,~V_\nu+
  g_{\nu_3}\tilde \alpha),
\nonumber\\&&
\left<\chi\right>=V(1+g\tilde \alpha)\ .
\label{correction}
\end{eqnarray}
Here, the parameters, $g_{l_i}$, $g_{\nu_i}$ and $g$, are of 
${\cal O}(1)$  
when $\alpha_l \sim \alpha_\nu \sim \alpha$ and couplings in the 
superpotential are of ${\cal O}(1)$. 
With these VEVs, the mass matrices are modified as,
\begin{eqnarray}
M_l
=v_d\alpha_l 
\begin{pmatrix}y_e(1+g_{l_1}\tilde \alpha)   
& y_eg_{l_3}\tilde \alpha & y_eg_{l_2}\tilde \alpha \\ 
y_\mu g_{l_2}\tilde \alpha & y_\mu(1+g_{l_1}\tilde \alpha) &
y_\mu g_{l_3} \tilde \alpha   \\
y_\tau g_{l_3}\tilde \alpha &y_\tau g_{l_2}\tilde \alpha & 
y_\tau(1+g_{l_1}\tilde  \alpha ) \\
 \end{pmatrix},
\end{eqnarray}
for the charged leptons, and
\begin{eqnarray}
M_\nu
=
\frac{y_1v_u^2\tilde \alpha(1+g\tilde \alpha)}{\Lambda} 
\begin{pmatrix}1   & 0 & 0 \\ 
                   0    & 0   &1    \\
                   0  & 1  & 0   \\
 \end{pmatrix}
+\frac{y_2v_u^2\tilde \alpha }{3\Lambda} 
\begin{pmatrix}2+2g_{\nu_1}\tilde \alpha   & -1-g_{\nu_3}\tilde \alpha
& -1-g_{\nu_2}\tilde \alpha \\ -1-g_{\nu_3}\tilde \alpha    & 
2+2g_{\nu_2}\tilde \alpha 
& -1-g_{\nu_1}\tilde \alpha  \\ -1-g_{\nu_2}\tilde \alpha  &
-1-g_{\nu_1}\tilde \alpha  & 2+2g_{\nu_3}\tilde \alpha \\
 \end{pmatrix} ,
\end{eqnarray}
for neutrinos.
These modified mass matrices give the deviation from the tri-bimaximal mixing
\footnote{Numerical analyses of the deviation from the tri-bimaximal mixing
 were presented in ref. \cite{honda}}.
That changes the lepton mixing angles by ${\cal O}(\tilde \alpha)$.
That implies that a large value of $\tan \beta$, $\tan \beta \gg 1$
and/or a small value of coupling $|y_\tau|$, $|y_\tau|\ll 1$ 
are unfavored in this scenario.
On the other hand, if $\tilde \alpha < {\cal O}(0.1)$, 
the above deviations in the lepton mass matrices 
would not be important to the current accuracy 
of neutrino oscillation experiments.
However, such deviations, in particular the deviation from the 
diagonal form in the charged lepton mass matrix, are important from 
the viewpoint of supersymmetry breaking terms as shown in 
the next section.
For the later convenience, here we show the diagonalizing 
matrix of the charged lepton mass matrix, that is, 
we diagonalize the charged 
lepton mass matrix $M_l$ in Eq.~(\ref{charged}) by
the matrices $V_L$ and $V_R$,
\begin{eqnarray}
\begin{split}
M_l=& V_R^T (\theta_{R12},\theta_{R23},\theta_{R13})
\begin{pmatrix}
m_e & 0 & 0 \\
0 & m_\mu & 0 \\
0 & 0 & m_\tau
\end{pmatrix} 
 V_L (\theta_{L12},\theta_{L23},\theta_{L13}) .
\end{split}
\label{OLR}
\end{eqnarray}
In the usual convention of $\theta_{R(L)ij}$, we get
\begin{eqnarray}
&& \theta_{R12}\sim \frac{y_e\langle\chi_3\rangle}{y_\mu\langle\chi_1\rangle}
\sim \frac{m_e}{m_\mu}{\cal O}(\tilde \alpha), ~~ 
\theta_{R23}\sim  \frac{y_\mu\langle\chi_3\rangle}{y_\tau\langle\chi_1\rangle}
\sim \frac{m_\mu}{m_\tau}{\cal O}(\tilde \alpha), ~~ 
 \theta_{R13}\sim  \frac{y_e\langle\chi_2\rangle}{y_\tau\langle\chi_1\rangle}
\sim \frac{m_e}{m_\tau}{\cal O}(\tilde \alpha), \nonumber\\
&&\theta_{L12}\sim \frac{\langle\chi_2\rangle}{\langle\chi_1\rangle}
\sim {\cal O}(\tilde \alpha), ~~ 
\theta_{L23}\sim  \frac{\langle\chi_2\rangle}{\langle\chi_1\rangle}
\sim {\cal O}(\tilde \alpha), ~~
 \theta_{L13}\sim  \frac{\langle\chi_3\rangle}{\langle\chi_1\rangle}
\sim {\cal O}(\tilde \alpha), 
\end{eqnarray}
where we take 
$\langle\chi_2\rangle\sim \langle\chi_3\rangle\sim 
\tilde \alpha \langle\chi_1\rangle$ taking account of
 non-leading corrections  as given in Eq.(\ref{correction}).
Here we have assumed $\tilde \alpha <{\cal O}(0.1)$.
Values of $\theta_{Lij}$ are large compared with those of 
$\theta_{Rij}$.


\section{Soft SUSY breaking terms}

We study soft SUSY breaking terms, i.e. 
soft slepton masses and A-terms, which are predicted from 
the $A_4$ model discussed in the previous section.
We consider SUSY breaking within the framework of supergravity theory, 
where some moduli fields $Z$ and $\chi$, $\chi_i$ and $\chi'_i$ 
would have non-vanishing F-terms.\footnote{SUSY breaking might be
  mediated through the gauge mediation and anomaly mediation. 
They are flavor-blind.
Only if the gravity mediation has a sizable contribution 
with and without other SUSY breaking mediations, 
we would have a prediction of sfermion spectra from each flavor 
mechanism.}
F-terms are given as  
\begin{eqnarray}
F^{\Phi_k}= - e^{ \frac{K}{2M_p^2} } K^{\Phi_k \bar{I} } \left(
  \partial_{\bar{I}} \bar{W} + \frac{K_{\bar{I}}} {M_p^2} \bar{W} \right) ,
\label{eq:F-component}
\end{eqnarray}
where $K$ denotes the K\"ahler potential, $K_{\bar{I}J}$ denotes 
second derivatives by fields, 
i.e. $K_{\bar{I}J}={\partial}_{\bar{I}} \partial_J K$
and $K^{\bar{I}J}$ is its inverse. 
In general, the fields $\Phi_k$ in our notation include 
$A_4 \times Z_3$-singlet moduli fields $Z$ and $\chi,~\chi_i,~\chi'_i$.
Furthermore, VEVs of $F_{\Phi_k}/\Phi_k$  are estimated as 
$\langle F_{\Phi_k}/ \Phi_k \rangle = {\cal O}(m_{3/2})$, where
$m_{3/2}$ denotes the gravitino mass, which is obtained as 
$m_{3/2}= \langle e^{K/2M_p^2}W/M_p^2 \rangle$.

\subsection{Slepton mass matrices}

First let us study soft scalar masses.
Within the framework of supergravity theory,
soft scalar mass squared is obtained as \cite{Kaplunovsky:1993rd}
\begin{eqnarray}
m^2_{\bar{I}J} K_{{\bar{I}J}}= m_{3/2}^2K_{{\bar{I}J}} 
+ |F^{\Phi_k}|^2 \partial_{\Phi_k}  
\partial_{  \bar{\Phi_k} }  K_{\bar{I}J}-
|F^{\Phi_k}|^2 \partial_{\bar{\Phi_k}}  K_{\bar{I}L} \partial_{\Phi_k}  
K_{\bar{M} J} K^{L \bar{M}}.
\label{eq:scalar}
\end{eqnarray}

The flavor symmetry $A_4 \times Z_3$ requires the following form 
of K\"ahler potential for left-handed and right-handed leptons 
\begin{eqnarray}
K^{(0)}_{\rm matter}
&=&a(Z,Z^\dagger)(L_e^\dagger L_e+L_\mu^\dagger L_\mu+L_\tau^\dagger L_\tau)
\nonumber\\&&
+b_e(Z,Z^\dagger)R_e^\dagger R_e
+b_\mu(Z,Z^\dagger)R_\mu^\dagger R_\mu
+b_\tau(Z,Z^\dagger)R_\tau^\dagger R_\tau ~ ,
\end{eqnarray}  
at the leading order, 
where $a(Z,Z^\dagger)$ and $b_I(Z,Z^\dagger)$ for
$I = e,\mu,\tau$ are generic functions of moduli fields $Z$.
Then, using eq.~(\ref{eq:scalar}), 
the slepton mass squared matrices of left-handed and right-handed 
sleptons can be found to be
\begin{eqnarray}
m_L^2 =
\left(
  \begin{array}{ccc}
m_{L}^2   &  0 &  0 \\ 
0   & m_{L}^2  & 0  \\ 
0   &  0  & m_{L}^2   \\ 
\end{array} \right),
\qquad
m_R^2
 = 
\left(
  \begin{array}{ccc}
m_{R_1}^2   &  0 &  0 \\ 
0   &  m_{R_2}^2 & 0  \\ 
0   & 0   & m_{R_3}^2   \\ 
\end{array} \right),
\label{eq:soft-mass-1}
\end{eqnarray}
where all of $m_L$ and $m_{R_i}$ for $i=1,2,3$ would be of  ${\cal O}(m_{3/2})$. 
These forms would be obvious from the flavor symmetry 
$A_4$, that is, three families of left-handed leptons are the 
$A_4$ triplet, while 
right-handed leptons are $A_4$ singlets.
At any rate, it is the prediction of 
the $A_4$ model that 
three families of left-handed slepton masses are degenerate.

However, the flavor symmetry $A_4 \times Z_3$ is broken 
to derive the realistic lepton mass matrices 
and such breaking introduces corrections in the K\"ahler potential 
and the form of slepton masses.
Let us study such corrections in the K\"ahler potential.
Because $\chi'_i$ and $\chi$ have nontrivial $Z_3$ charges, 
their linear terms do not appear in the 
K\"ahler potential of lepton multiplets.
In addition, because of 
$\langle \chi_2 \rangle, \langle \chi_3 \rangle \sim 
\tilde \alpha \langle \chi_1 \rangle$, the 
most important correction terms would be linear terms of 
$\chi_1$.
That is, the correction terms in the matter K\"ahler potential 
are obtained 
\begin{equation}
\Delta K_{\rm matter} = \frac{\chi_1}{\Lambda'}\left[ a'_1(Z,Z^\dagger)
(2 L_e^\dagger L_e -L_\mu^\dagger L_\mu - L_\tau^\dagger L_\tau)  
 + a'_2(Z,Z^\dagger)(L_\mu^\dagger L_\mu - L_\tau^\dagger L_\tau)
\right ] 
+ h.c.,
\label{eq:K-corr}
\end{equation}
up to ${\cal O}(\tilde \alpha^2\Lambda/\Lambda')$, 
where $a'_1(Z,Z^\dagger)$ and $a'_2(Z,Z^\dagger)$ are generic 
functions of moduli fields.
The cut-off scale $\Lambda'$ might be independent of $\Lambda$, 
which appears in the effective superpotential.
For example, if $\Lambda'$ is the Planck scale, 
the above corrections would be negligible.
Hereafter, we concentrate to the case with 
$\Lambda' \sim \Lambda$.
Note that linear correction terms of $\chi_i$ do not appear 
for the K\"ahler potential of right-handed lepton multiplets.
All of off-diagonal K\"ahler metric entries for both 
left-handed and right-handed leptons appear at ${\cal O}(\tilde
\alpha^2)$,
\begin{equation}
\frac{\partial^2 K_{\rm matter}}{\partial L^\dagger_{\bar I} 
\partial L_J} ={\cal O}(\tilde \alpha^2), \qquad 
\frac{\partial^2 K_{\rm matter}}{\partial R^\dagger_{\bar I} 
\partial R_J} ={\cal O}(\tilde \alpha^2),
\label{eq:K-off-diag}
\end{equation}
where $I,J=e, \mu, \tau$ and $I \neq J$.
For example, the (1,2) and (2,1) entries for left-handed leptons 
are induced by $(\chi_3/\Lambda')L^\dagger_e L_\mu$, 
$(\chi_2/\Lambda')L^\dagger_\mu L_e$, etc.
Similarly, other entries for both left-handed and right-handed 
leptons are induced.
Furthermore, corrections including $\Phi$ do not violate the 
structure of $K^{(0)}_{\rm matter}$ because 
$\Phi$ has a trivial charge under $A_4 \times Z_3$.

Including these corrections, 
the slepton masses are written by 
\begin{eqnarray}
m_L^2&=&
\left(
  \begin{array}{ccc}
m_{L}^2   &  0 &  0 \\ 
0   & m_{L}^2  & 0  \\ 
0   &  0  & m_{L}^2   \\ 
\end{array} \right)
+
m_{3/2}^2
\left(
  \begin{array}{ccc}
2\bar a'_1\tilde \alpha   &  0 & 0 \\ 
0   &  (\bar a'_2-\bar a'_1)\tilde \alpha  & 0 \\ 
0   &  0 & -(\bar a'_1+ \bar a'_2)\tilde \alpha  \\ 
\end{array} \right) + {\cal O}(\tilde \alpha ^2m_{3/2}^2),
\nonumber\\
m_R^2
&=&
\left(
  \begin{array}{ccc}
m_{R_1}^2   &  0 &  0 \\ 
0   &  m_{R_2}^2 & 0  \\ 
0   & 0   & m_{R_3}^2   \\ 
\end{array} \right)
+{\cal O}(\tilde \alpha ^2m_{3/2}^2),
\label{eq:soft-mass-2}
\end{eqnarray}
in the flavor basis, 
where $\bar a'_1, \bar a'_2 ={\cal O}(1)$.

The leptonic FCNC is induced by  off diagonal elements of scalar 
mass squared matrices 
in the diagonal basis of the charged lepton mass matrix.
The following discussion presents that the off diagonal elements
are  enough suppressed  in the left-handed slepton and the right-handed
slepton mass matrices 
in the diagonal basis of the charged lepton mass matrix, 
i.e., $\tilde m^2_{R}=V_R m_R^2 V_R^T$ and  
$\tilde m^2_{L}=V_L m_L^2 V_L^T$.
In this basis, the slepton mass squared matrices 
are obtained as 
\begin{eqnarray}
\tilde m_L^2&=&
\left(
  \begin{array}{ccc}
m_{L}^2   &  0 &  0 \\ 
0   & m_{L}^2  & 0  \\ 
0   &  0  & m_{L}^2   \\ 
\end{array} \right)
+
m_{3/2}^2
\left(
  \begin{array}{ccc}
 O(\tilde \alpha)   &  O(\tilde \alpha ^2) & O(\tilde \alpha ^2) \\ 
O(\tilde \alpha ^2)   &  O(\tilde \alpha)  & O(\tilde \alpha ^2) \\ 
O(\tilde \alpha ^2)   &  O(\tilde \alpha ^2)& O(\tilde \alpha)   \\ 
\end{array} \right),
\nonumber\\
\tilde m_R^2
&=&
\left(
  \begin{array}{ccc}
m_{R_1}^2   &  0 &  0 \\ 
0   &  m_{R_2}^2 & 0  \\ 
0   & 0   & m_{R_3}^2   \\ 
\end{array} \right)
+{\cal O}(\tilde \alpha ^2m_{3/2}^2) ,
\label{eq:soft-mass-SCKM}
\end{eqnarray}
when $\tilde \alpha >  {m_e}/{m_\mu}$.

We have a strong constraint on $(\tilde m^2_{L})_{12}$ and 
$(\tilde m^2_{R})_{12}$ from FCNC experiments~\cite{FCNCbound}, i.e. 
\begin{equation}
\frac{(\tilde m^2_{L})_{12}}{m^2_{\rm SUSY}}\leq 
{\cal O}(10^{-3}), \qquad 
\frac{(\tilde m^2_{R})_{12}}{m^2_{\rm SUSY}} \leq 
{\cal O}(10^{-3}) ,
\end{equation}
for $m_{\rm SUSY} \sim 100$ GeV, 
where $m_{\rm SUSY}$ denotes the average mass of slepton masses and 
it would be of ${\cal O}(m_{3/2})$.
The above prediction (\ref{eq:soft-mass-SCKM}) 
of the $A_4$ model leads to 
$(\tilde m^2_{L})_{12}/m^2_{\rm SUSY} ={\cal O}(\tilde \alpha^2)$.
Because of $\tilde \alpha \sim 0.03$ for $y_\tau \simeq 1$, 
our prediction,  
$(\tilde m^2_{L})_{12}/m^2_{\rm SUSY} ={\cal O}(\tilde \alpha^2) =
{\cal O}(10^{-3})$, would be consistent with the current experimental 
bound.
When we consider a larger value of $y_\tau$, e.g. 
$y_\tau \sim 3$, the predicted value of $(\tilde
m^2_{L})_{12}/m^2_{\rm SUSY}$ would be suppressed like 
$(\tilde m^2_{L})_{12}/m^2_{\rm SUSY} ={\cal O}(10^{-4})$.
On the other hand, a large value of $\tilde \alpha$ like 
$\tilde \alpha ={\cal O}(0.1)$, which 
is obtained for a large value of $\tan \beta$ and/or 
a small value of $y_\tau$ would be ruled out.
Similarly, we can estimate $(\tilde m^2_{R})_{12}/m^2_{\rm SUSY}$
by using eq.~(\ref{eq:soft-mass-SCKM}) and 
results are the same.

We have studied soft scalar masses induced by F-terms.
If we gauge $U(1)_F$, another contribution to scalar masses 
would be induced through $U(1)_F$ breaking, that is, 
contributions due to the D-term of the $U(1)_F$ vector 
multiplet.
Such D-term contributions $m_D^2$ are proportional to 
$U(1)_F$ charges $Q$ of matter fields~\footnote{See e.g. 
\cite{Kawamura:1996wn}.}, 
\begin{equation}
m_D^2=Q\langle D
\rangle.
\end{equation}
In general, such D-term contributions may be dangerous from 
the viewpoint of FCNC.
However, those $D$-term contributions in the $A_4$ model 
do not violate the above 
form of soft scalar masses, (\ref{eq:soft-mass-1}),
(\ref{eq:soft-mass-2}) and (\ref{eq:soft-mass-SCKM}), 
because $U(1)_F$ charges in 
Table 1 are consistent with the $A_4$ flavor symmetry, 
that is, $L_I$ $(I=e, \mu, \tau)$ have the same $U(1)_F$ charge, while 
three of $R_I$ $(I=e, \mu, \tau)$ have different $U(1)_F$ charges.
Thus, the predictions on $(\tilde m^2_{L})_{12}/m^2_{\rm SUSY}$ 
and $(\tilde m^2_{R})_{12}/m^2_{\rm SUSY}$ do not change.

Here, we give a comment on radiative corrections.
The slepton masses, which we have studied above, are 
induced at a high energy scale such as the Planck scale 
or the GUT scale.
The slepton masses have radiative corrections between 
such a high energy scale and the weak scale, 
although those have been neglected in the above analyses.
In those radiative corrections to slepton masses, 
the gaugino contributions are dominant.
For example, slepton masses at the weak scale 
are related with ones at the GUT scale $M_X$ as 
\begin{eqnarray}
m_{L}^2(M_Z) &= & m_{L}^2(M_X) + 0.5 M_{\tilde W}^2 
+ 0.04M_{\tilde B}^2, 
\nonumber \\
m_{R}^2(M_Z) &= & m_{R}^2(M_X) + 0.2 M_{\tilde B}^2, 
\end{eqnarray}
where $M_{\tilde B}$ and $M_{\tilde W}$ are bino and wino masses, 
respectively.
These radiative corrections do not change drastically 
the above results when these gaugino masses are of 
${\cal  O}(m_{3/2})$.
On the other hand, 
FCNC constraints would be improved when 
these gaugino masses are much larger than 
initial values of slepton masses.

\subsection{A-terms}

Now, let us examine the mass matrix between the left-handed and
the right-handed sleptons, which is generated by the so-called A-terms.
The A-terms are trilinear couplings of two sleptons and one Higgs field, 
and are obtained as \cite{Kaplunovsky:1993rd}
\begin{equation}
h_{IJ} {L}_J {R}_I H_d =  
h^{(Y)}_{IJ}{L}_J {R}_I H_d  + h^{(K)}_{IJ}{L}_J {R}_I H_d ,
\label{eq:A-term}
\end{equation}
where 
\begin{eqnarray}
h^{(Y)}_{IJ} &=& F^{\Phi_k} \langle \partial_{\Phi_k} \tilde{y}_{IJ}
\rangle ,  
\nonumber \\
h^{(K)}_{IJ}{L}_J {R}_I H_d &=& - 
\langle \tilde{y}_{LJ} \rangle {L}_J {R}_I H_d F^{\Phi_k} K^{L\bar{L}}
\partial_{\Phi_k} K_{\bar{L}I}  \\
& &  -
\langle \tilde{y}_{IM} \rangle {L}_J {R}_I H_d F^{\Phi_k} K^{M\bar{M}} 
\partial_{\Phi_k} K_{\bar{M}J}  \nonumber  \\ 
& & -   \langle \tilde{y}_{IJ} \rangle {L}_J {R}_I H_d F^{\Phi_k} K^{H_d}
\partial_{\Phi_k} K_{H_d}, \nonumber  
\label{eq:A-term-2}
\end{eqnarray}
where $K_{H_d}$ denotes the K\"ahler metric of $H_d$,
and $\tilde{y}_{IJ}$ denotes the effective Yukawa couplings 
given as 
\begin{eqnarray}
\tilde{y}_{IJ}
&=&\frac{1}{\Lambda} 
\left(
  \begin{array}{ccc}
y_e  \chi_1&  y_e  \chi_3 &   
y_e \chi_2 \\ 
 y_\mu  \chi_2  &  y_\mu  \chi_1 &  
y_\mu  \chi_3 \\ 
 y_\tau  \chi_3 & y_\tau  \chi_2 & 
 y_\tau \chi_1 \\ 
\end{array} \right) \ .
\label{yukawal}
\end{eqnarray}
Furthermore, when we use the $U(1)_F$ Froggatt-Nielsen mechanism 
in order to obtain the lepton mass hierarchy, the couplings, 
$y_e, y_\mu$ and $
y_\tau$, are expressed as 
\begin{eqnarray}
y_{I}=c_{I} \left( \frac{\Phi}{\Lambda} \right)^{Q_I} ~~(I
=e, \mu, \tau) ,
\end{eqnarray}
where $Q_I$ is $U(1)_F$ charges.
Here we assume that couplings, $c_e,c_\mu$ and $c_\tau$, do not 
include the moduli $Z$, i.e. $\partial_Z c_I =0$ for $I =e, \mu , \tau$.
After the electroweak symmetry breaking, these $A$-terms provide us 
with the left-right mixing mass squared $(m^2_{LR})_{IJ} = h_{IJ}v_d$.
Furthermore, we use the basis
$(\tilde m^2_{LR})_{IJ} = (V_R m^2_{LR} V_L^T)_{IJ}$.
The third term in the right hand side of eq.(\ref{eq:A-term-2}) 
is diagonalized in this basis.
Thus, we do not take the third term into account 
in the following discussion.

When we consider the leading order of K\"ahler potential 
$K^{(0)}_{\rm  matter}$, 
the second terms in the right hand side of eq.~(\ref{eq:A-term-2}), 
$h^{(K)}_{IJ}$, 
is written by \cite{Kobayashi:2000br}
\begin{equation}
h^{(K)}_{IJ} = \langle \tilde y_{IJ} \rangle (A^R_I+A^L_J),
\label{eq:h-K}
\end{equation}
where $A^R_I = -F^Z\partial_Z \ln b_I(Z,Z^\dagger) $ 
and $A ^L_J =  -F^Z\partial_Z \ln a(Z,Z^\dagger) $, 
that is, $A^L_I$ are degenerate up to ${\cal O}(\tilde \alpha)$.
Thus, the (2,1) entry of $(\tilde m^2_{LR})_{IJ}$ vanishes 
at the leading order.
However, such a behavior is violated at the next order, that is, 
$A ^L_1-A^L_2 = {\cal O}(\tilde \alpha m_{3/2})$, because 
the diagonal (1,1) and (2,2) entries of K\"ahler metric $\Delta
K_{matter}$ 
for the left-handed lepton multiplets (\ref{eq:K-corr}) 
have non-degenerate corrections of 
${\cal  O}(\tilde \alpha)$.
Then, the $h^{(K)}_{IJ}$ contribution to 
the (2,1) entry of $(\tilde m^2_{LR})_{IJ}$ is estimated as 
\begin{equation}
(\tilde m^2_{LR})_{21} \sim \langle \tilde y_\mu \rangle 
v^d (A ^L_1-A^L_2)\theta_{L12} = 
{\cal O}(\tilde \alpha^2 m_\mu m_{3/2}).
\end{equation}
Furthermore, the off-diagonal elements of K\"ahler metric 
have ${\cal O}(\tilde \alpha^2)$ of corrections (\ref{eq:K-off-diag}), 
and these 
corrections also induce the same order of $(\tilde m^2_{LR})_{21}$, i.e. 
$(\tilde m^2_{LR})_{21} = 
{\cal O}(\tilde \alpha^2 m_\mu m_{3/2})$.
Similarly, we can estimate the (1,2) entry and obtain 
the same result, i.e., $(\tilde m^2_{LR})_{12} = 
{\cal O}(\tilde \alpha^2 m_\mu m_{3/2})$
when $\tilde \alpha >  {m_e}/{m_\mu}$.
These entries have the strong constraint from FCNC experiments as 
$(\tilde m^2_{LR})_{12}/m^2_{\rm SUSY} \leq {\cal O}(10^{-6})$
and the same for the (2,1) entry for $m_{\rm SUSY} =100$ GeV.
However, the above prediction of the $A_4$ model leads to 
$(\tilde m^2_{LR})_{12}/m^2_{\rm SUSY} = {\cal O}(10^{-7})$ 
for $m_{\rm SUSY} =100$ GeV and $\alpha \sim 0.03$ 
and that is consistent with the experimental bound.

Now, let us estimate the first term in the right hand side  
of (\ref{eq:A-term}), $h^{(Y)}_{IJ}$, which can be written by, 
\begin{eqnarray}
\begin{split}
(h^{(Y)}_{IJ} )=& \left( \frac{1}{\Lambda} \right)
\begin{pmatrix} 
y_e \langle \chi_1 \rangle A_1  & y_e \langle \chi_3 \rangle A_3  
& y_e \langle \chi_2 \rangle A_2   \\ 
y_{\mu} \langle \chi_2 \rangle A_2   & y_{\mu} \langle \chi_1  \rangle
A_1  
  & y_{\mu} \langle \chi_3 \rangle A_3 \\ 
y_{\tau} \langle \chi_3 \rangle A_3 & y_{\tau} \langle \chi_2 \rangle
A_2 
& y_{\tau} \langle \chi_1 \rangle A_1 
\end{pmatrix} \\
&+\left( \frac{A_0}{\Lambda} \right) \begin{pmatrix} Q_e & 0 & 0 \\ 0
  &  
Q_{\mu} & 0 \\ 0 & 0 & Q_{\tau} \end{pmatrix}
\begin{pmatrix} 
y_e \langle \chi_1 \rangle  & y_e \langle \chi_3 \rangle  & 
y_e \langle \chi_2  \rangle  \\ 
y_{\mu} \langle \chi_2 \rangle  & y_{\mu} \langle \chi_1 \rangle      
& y_{\mu} \langle \chi_3  \rangle \\ 
y_{\tau} \langle \chi_3  \rangle & y_{\tau} \langle \chi_2  \rangle 
& y_{\tau} \langle \chi_1 \rangle 
\end{pmatrix} , \\
\end{split}
\label{eq:h-Y}
\end{eqnarray}
where  
\begin{eqnarray}
 A_0\equiv \frac{F_{\Phi}}{\Phi}, \qquad  
A_i \equiv \frac{F_{\chi_i}}{\chi_i}\ ,  \quad (i=1,2,3).
\end{eqnarray}
These would be of ${\cal O}(m_{3/2})$.
Since the second term of (\ref{eq:h-Y}) is exactly the form 
of eq.~(\ref{eq:h-K}) with the degenerate $A^L_I$, 
the second term does not change 
the above estimation of 
$(\tilde m^2_{LR})_{12}$ and $(\tilde m^2_{LR})_{21}$.\footnote{
Indeed, the K\"ahler metric $b_I(Z,Z^\dagger)$ and couplings 
$y_I = c_I (\Phi/\Lambda)^{Q_I}$ lead to the same 
A-terms as the K\"ahler metric $b_I(Z,Z^\dagger)(\Phi/\Lambda)^{-Q_I}$
and couplings $y_I = c_I$.}

The first term of (\ref{eq:h-Y}) contributes to 
$(\tilde m^2_{LR})_{21}$ as 
\begin{equation}
(\tilde m^2_{LR})_{21} = y_\mu v_d\chi_2(A_2 -A_1)/\Lambda 
\sim m_\mu \tilde \alpha  (A_2-A_1).
\end{equation}
That is, we estimate 
$(\tilde m^2_{LR})_{21}/m^2_{\rm SUSY} \sim 10^{-5} \times
(A_2-A_1)/m_{3/2}$ for $\tilde \alpha \sim 0.03$.
Thus, if $A_2 \neq A_1$ and $A_i ={\cal O}(m_{3/2})$, 
this value of  $(\tilde m^2_{LR})_{21}/m^2_{\rm SUSY}$ 
would not be consistent with the experimental bound 
for $m_{\rm SUSY} =100$ GeV.
Hence, a smaller value of $\tilde \alpha$ like 
$\tilde \alpha ={\cal O}(0.001)$ would 
be favorable to be consistent with the experimental bound and 
that implies $y_\tau \sim 4 \pi$.
Alternatively, for  $\tilde \alpha \sim 0.03$ 
it is required that $A_1 = A_2$ up to ${\cal O}(0.1)$.
If the non-trivial superpotential leading to 
SUSY breaking does not include $\chi_i$, i.e. 
$\langle \partial_{\chi_i} W \rangle =0$, we can realize 
\begin{equation}
A_i = -  \left( K^{(\chi)}_{i\bar i}\right)^{-1}m_{3/2},
\end{equation}
where $K^{(\chi)}_{i\bar i}$ is 
the K\"ahler metric of the fields $\chi_i$.
Note that the K\"ahler metric for $\chi_i$ are degenerate 
at the leading order, because $\chi_i$ are the $A_4$ triplet.
Hence, we obtain the degeneracy between $A_i$, i.e., 
$A_1=A_2=A_3$ up to ${\cal O}(\tilde \alpha m_{3/2})$.
In this case, $(\tilde m^2_{LR})_{21}$ is suppressed 
and we can estimate 
$(\tilde m^2_{LR})_{21}/m^2_{\rm SUSY} \sim \tilde \alpha^2 m_\mu/m_{3/2}
={\cal O}(10^{-6})$ for $\tilde \alpha \sim 0.03$. 
This value is consistent with the experimental bound.
However, the parameter region with larger $\alpha$ like 
$\tilde \alpha = {\cal O}(0.1)$ is still ruled out.
Obviously, the value of $(\tilde m^2_{LR})_{21}$ depends 
on the difference between $A_1$ and $A_2$.
If the difference $A_1 -A_2$ is smaller than ${\cal O}(\tilde \alpha
m_{3/2})$, $(\tilde m^2_{LR})_{21}$ would be suppressed more.

Here we give a comment on radiative corrections.
Similarly to slepton masses, radiative corrections 
to A-terms do not change drastically the above results.
Note that Yukawa couplings are small, in particular 
the first and second families.

\subsection{Comparison with other models}

Here we give briefly comments 
comparing the $A_4$ model and other models.
First, let us compare with the $D_4$
models~\cite{Grimus,Ishimori:2008gp,Ishimori:2008ns}, which 
also lead to the tri-bimaximal mixing for the lepton sector 
by choosing proper values of parameters.
In both of the $A_4$ model and the $D_4$ models, 
the charged lepton mass matrices, $M_l$,  
are diagonal at the leading order, but 
there are small corrections at the next order.
That is, in the $A_4$ model, the (1,2) entries, $\theta_{L12}$ 
and $\theta_{R12}$, of diagonalizing matrices of $M_l$ 
are estimated as  $\theta_{L12} = {\cal O}(\tilde \alpha)$ and 
 $\theta_{R12} = {\cal O}(\tilde \alpha m_e/m_\mu)$ 
with $\tilde \alpha \sim 0.03$ for the typical values, 
while in the $D_4$ models, both of those entries are estimated as 
$\theta_{L12} = \theta_{R12} = {\cal O}(10^{-2})$ - ${\cal
  O}(10^{-3})$.
Thus, these angles would be smaller in the $D_4$ models.

In the $D_4$ models, 
the second and third families of left-handed and right-handed 
charged leptons correspond to $D_4$ doublets 
and the first families correspond to $D_4$ singlets.
Then,  the second and third families of 
left-handed and right-handed slepton masses are degenerate 
at the leading order, while the first families of 
sleptons masses are, in general, independent of the others.
On the other hand, in the $A_4$ model, all the three families 
of left-handed slepton masses are degenerate 
at the leading order,  while 
the three families of right-handed slepton masses are,
in general, different from each other.
That is, we have different predictions in 
slepton mass spectra between the $A_4$ model and the $D_4$ models.

In the $A_4$ model, a large value of 
the entry $\theta_{L12}={\cal O}(\tilde \alpha)$  
like $\theta_{L12} \sim \tilde \alpha \sim 0.03$ 
would not be favorable from the viewpoint of 
FCNC experimental bounds.
However, the triplet structure of 
left-handed leptons is helpful 
to suppress $(\tilde m^2_{LL})_{12}$ and $(\tilde m^2_{LR})_{12}$ 
and the predicted values in a wide parameter space 
become consistent with the FCNC experimental bounds.

On the other hand, in the $D_4$ models, 
small values of mixing angles 
$\theta_{L12} = \theta_{R12} = {\cal O}(10^{-3})$ 
are helpful to suppress $(\tilde m^2_{LL})_{12}$, 
$(\tilde m^2_{RR})_{12}$ and $(\tilde m^2_{LR})_{12}$, 
although the first and second families of slepton masses 
are not degenerate.
This situation is the same as one for the right-handed 
sleptons in the $A_4$ model.
Then, both models are consistent with 
the current FCNC experimental bounds.

Next, we give a comment on other $A_4$ models.
Indeed, several $A_4$ models have been proposed.
For those SUSY $A_4$ models, we can evaluate 
soft SUSY breaking terms in a way similar to 
section 3.1 and 3.2.
We would obtain similar results in the models that 
left-handed and right-handed leptons are assigned 
to a triplet and singlets, respectively, and 
the angles, $\theta_{L12}$ and $\theta_{R12}$, are 
similar to the above values.
Alternatively, we could construct the supersymmetric model, where 
three families of right-handed leptons are assigned 
with an $A_4$ triplet, while three families of left-handed 
leptons are assigned with three singlets, 
${\bf  1}$, ${\bf  1}'$ and ${\bf  1}''$, 
that is the opposite assignment of the $A_4$ model
of \cite{Alta2}.\footnote{Indeed, such non-SUSY model has been 
studied in~\cite{kaneko}.}
In such a model, three families of right-handed slepton masses 
would be degenerate at the leading order, 
since the right-handed leptons are a triplet.
On the other hand, three families of left-handed slepton masses 
would be different from each other.
Thus, the prediction on soft SUSY breaking terms 
depend on flavor symmetries and assignments of 
matter fields.

\section{Conclusion}

We have studied soft SUSY breaking terms, which are 
derived from the $A_4$ model.
Three families of left-handed slepton masses are degenerate, 
while three families of right-handed slepton masses 
are, in general, different from each other.
That is the pattern different from slepton masses 
in the $D_4$ model, where only the second 
and third families of both left-handed and right-handed 
slepton masses are degenerate~\cite{Ishimori:2008ns}.

In the wide parameter region, the FCNCs 
predicted in the SUSY $A_4$ model are consistent with 
the current experimental bounds.
Thus, the non-Abelain flavor symmetry 
in the $A_4$ model is useful not only to derive 
realistic lepton mass matrices, but also to 
suppress FCNC processes.
If the bound of $BR(\mu \rightarrow e \gamma)$ 
is improved in future, e.g. by the MEG experiment 
\cite{Mori:2007zza}, the allowed parameter space 
would be reduced, that is, a smaller value of $\tilde \alpha$, 
i.e. a larger value of $y_\tau$ like $y_\tau\sim 4 \pi$, 
would become favorable.

\vskip 1cm
{\bf Note to be added}

While this paper was being completed, Ref.~\cite{Feruglio:2008ht} 
appeared, where a similar issue was studied.

\vspace{1cm}
\noindent
{\bf Acknowledgement}

T.~K. and Y.~O. are supported in part by the
Grant-in-Aid for Scientific Research, No. 20540266 
and No. $20\cdot 324$, and 
the Grant-in-Aid for the Global COE Program 
"The Next Generation of Physics, Spun from Universality 
and Emergence" from the Ministry of Education, Culture,
Sports, Science and Technology of Japan.
The work of M.T. has been  supported by the
Grant-in-Aid for Science Research
of the Ministry of Education, Science, and Culture of Japan No. 17540243.


\section*{Appendix}
\appendix
\section{$A_4$ group}

The group $A_4$ has two generators $S$ and $T$, which satisfies 
$S^2=(ST)^3=T^3=1$.
In the representation, where 
$T$ is taken  to be diagonal, the elements $S$ and $T$ are expressed as
\begin{eqnarray}
S=   \begin{pmatrix}-1   & 2 & 2 \\ 
                   2    & -1   & 2    \\
                   2   & 2  & -1     \\
 \end{pmatrix},\quad
T=   \begin{pmatrix}1   & 0 & 0 \\ 
                   0    & \omega^2   &0    \\
                   0  & 0  & \omega    \\
 \end{pmatrix}. 
\end{eqnarray}
Then, twelve elements of  the $A_4$ group are given by 
\begin{eqnarray}
1,\quad
S,\quad
T,\quad
ST,\quad
TS,\quad
T^2,\quad
\nonumber\\
ST^2,\quad
STS,\quad
TST,\quad
T^2S,\quad
TST^2,\quad
T^2ST.
\end{eqnarray}

The product of two triplets $\bf 3\times 3$, where
\begin{eqnarray}
a=(a_1,a_2,a_3),
\quad
b=(b_1,b_2,b_3).
\end{eqnarray}
 is decomposed as 
\begin{equation}
{\bf 3} \times {\bf 3} = {\bf 1} + {\bf 1}' + {\bf 1}'' + {\bf 3} +
{\bf 3}',
\end{equation}
where
\begin{eqnarray}
 & & {\bf 1}  : (a_1b_1+a_2b_3+a_3b_2), \quad {\bf 1}':
(a_3b_3+a_1b_2+a_2b_1), \quad {\bf 1}'':
(a_2b_2+a_1b_3+a_3b_1) ,
\nonumber\\
 & & {\bf 3} : 
(2a_1b_1-a_2b_3-a_3b_2,2a_3b_3-a_1b_2-a_2b_1,2a_2b_2-a_1b_3-a_3b_1), 
\\
 & & {\bf 3}':(a_2b_3-a_3b_2,a_1b_2-a_2b_1,a_1b_3-a_3b_1). 
\nonumber
\end{eqnarray}

\end{document}